\pdfoutput=1
\documentclass{aastex63}

\usepackage{amsmath}
\usepackage{xfrac}
\accepted{September 14, 2022}
\begin{document}

\title{Kinematical Constraint on Eccentricity in the Protoplanetary Disk MWC~758 with ALMA}

\author{Kuo, I-Hsuan Genevieve}
\affiliation{Academia Sinica Institute of Astronomy and Astrophysics, 11F of Astro-Math Bldg, 1, Sec. 4, Roosevelt Rd, Taipei 10617, Taiwan}
\affiliation{Department of Physics, National Taiwan University, No. 1, Section 4, Roosevelt Rd., Taipei 10617, Taiwan}

\author{Yen, Hsi-Wei}
\affiliation{Academia Sinica Institute of Astronomy and Astrophysics, 11F of Astro-Math Bldg, 1, Sec. 4, Roosevelt Rd, Taipei 10617, Taiwan}

\author{Gu, Pin-Gao}
\affiliation{Academia Sinica Institute of Astronomy and Astrophysics, 11F of Astro-Math Bldg, 1, Sec. 4, Roosevelt Rd, Taipei 10617, Taiwan}

\author{Chang, Tze-En}
\affiliation{Academia Sinica Institute of Astronomy and Astrophysics, 11F of Astro-Math Bldg, 1, Sec. 4, Roosevelt Rd, Taipei 10617, Taiwan}
\affiliation{Department of Physics, Imperial College London, London SW7 2AZ, UK.}


\begin{abstract}
We analyzed the archival data of the $^{13}\mathrm{CO}$ and $\mathrm{C}^{18}\mathrm{O}$ $J=3-2$ emission lines in the protoplanetary disk around MWC~758 obtained with the Atacama Large Millimeter/submillimeter Array to discuss possible planet-disk interaction and non-Keplerian motion in the disk. We performed fitting of a Keplerian disk model to the observational data and measured the velocity deviations from the Keplerian rotation. We found significant velocity deviations around the inner cavity in the MWC~758 disk. We examined several possibilities that may cause the velocity deviations, such as pressure gradient, height of the emitting layer, infall motion, inner warp, and eccentricity in the disk. We found that the combination of an eccentric orbital motion with eccentricity of $0.1\pm0.04$ at the radius of the inner cavity and an infalling flow best explains the observed velocity deviations. Our kinematically constrained eccentricity of the gas orbital motion close to the inner cavity is consistent with the eccentricity of the dust ring around the inner cavity measured in the submillimeter continuum emission. Our results hint at strong dust-gas coupling around the inner cavity and presence of a gas giant planet inside the inner cavity in the MWC~758 disk.

\end{abstract}

\section{Introduction} \label{sec:intro}

Planets form in protoplanetary disks, which could induce various substructures such as multiple dust rings, gaps, cavities, and clumps \citep{DSHARP_VII}, spirals \citep{Ogilvie_2002}, or eccentric orbits \citep{Kley_2006} in the disks. These substructures may in turn introduce instability or pressure maxima that may lead to further planet formation \citep{Pinilla_2015}. The small size and low thermal temperature of protoplanets, along with shielding from abundant gas and dust in protoplanetary disks, pose difficulties in direct imaging of protoplanets. Therefore, the substructures in protoplanetary disks are important observables that can provide clues to the process of planet formation.
\par
The complex morphology of the protoplanetary disk of MWC~758 serves as a suitable site to study disk substructures and probable planet-disk interaction. MWC~758 is an Herbig Ae star located at a distance of $160\pm2$ pc \citep{gaia_dr2}, with an age of $3.5\pm2$ Myr \citep{Meeus_2012}. The most current models have deduced its spectroscopic mass to be between 1.5 and 1.9 $M_\sun$ \citep{Vioque_2018, Garufi_2018}. Numerous previous studies in near-infrared and centimeter to millimeter/submillimeter wavelengths have revealed dust clumps \citep{Casassus_2019}, spiral arms \citep{Grady_2013, Benisty_2015}, and an inner cavity with a radius up to $\sim50$ AU \citep{Andrews_2011}. One unique structure is an eccentric inner dust ring with an eccentricity of $0.1\pm0.01$ at a radius of $0\farcs32$ detected in the 0.88 mm continuum emission with the Atacama Large Millimeter/submillimeter Array (ALMA) observations \citep{Dong_2018}.
\par
Disk eccentricity can be excited via the 3:1 eccentric Lindblad resonance and undergo exponential growth, if a secondary potential is present in a star-disk system and an initial dynamical perturbation is induced \citep{Lubow_1991}. Eccentric Lindblad resonances due to planet-disk tidal interactions are also responsible for eccentricity excitation of protoplanetary disks \citep[e.g.][]{Goldreich_Tremaine_1981, Teyssandier_Ogilvie_2016}. Previous studies involving hydrodynamic simulations have found that for a typical $\alpha$ viscosity of $10^{-5}$ in a protoplanetary disk, a planet-star mass ratio of 
$\sim10^{-2}-10^{-3}$ can generate disk eccentricities of 
$\sim0.1-0.2$ \citep{Kley_2006, Hsieh_2012, Ataiee_2013, Teyssandier_Ogilvie_2017}. In addition, the gas motion in simulations closely satisfies the eccentric equation at $3a_p\ge r\ge2a_p$, where $a_p$ is the orbital radius of the planet, likely due to the effect of the irregular horseshoe orbit and nonlinear dynamics in the vicinity of the planetary potential, and the eccentricity profile undergoes rapid decay when $r>3a_p$. The dust grains can also move on an elliptical orbit in an eccentric gaseous disk arising from the aerodynamical drag as well as the secular gravitational potential of the embedded planet \citep{Hsieh_2012}.
\par
Therefore, from theory we expect that the gas rotation close to the inner cavity, where the dust distribution is eccentric, in the MWC~758 disk is also eccentric \citep{Gu_2014}, 
while the eccentricity of the gas rotation in MWC~758 has not yet been constrained.
In this work, we use ALMA archival data of $^{13}\mathrm{CO}$ $J=3-2$ and $\mathrm{C}^{18}\mathrm{O}$ $J=3-2$ at an angular resolution of $0\farcs26\times0\farcs17$, and radiative transfer modelling to constrain the gas eccentricity in the MWC~758 disk from the gas kinematics.

\section{Data} \label{sec:data}

ALMA Cycle 3 Band 7 visibility data of the $^{13}\rm{CO}$ $J=3-2$, $\rm{C}^{18}\rm{O}$  $J=3-2$, and 0.88 mm continuum emission in MWC~758 (project code: 2012.1.00725.S) were retrieved from the public ALMA archive. Details of the observation have been described in \citet{Boehler_2018}. We calibrated the raw visibility data with the script obtained from the archive using the pipeline of the Common Astronomy Software Applications (CASA) of version 4.3.1.  The continuum was subtracted from the $^{13}\rm{CO}$ $J=3-2$ and $\rm{C}^{18}\rm{O}$  $J=3-2$ line data with the CASA task \textit{uvcontsub}. Then, the $^{13}\rm{CO}$ $J=3-2$, $\rm{C}^{18}\rm{O}$  $J=3-2$, and continuum images were generated with the CASA task \textit{tclean}. Superuniform weighting was adopted for the continuum image, while natural weighting was adopted for the line images to optimize the signal-to-noise ratio (S/N). The channel width of the line images was adopted to be 0.1 km\,s$^{-1}$. A summary of the images is given in Table \ref{tab:images}. 

\begin{figure*}[ht!]
\epsscale{1.15}
\plotone{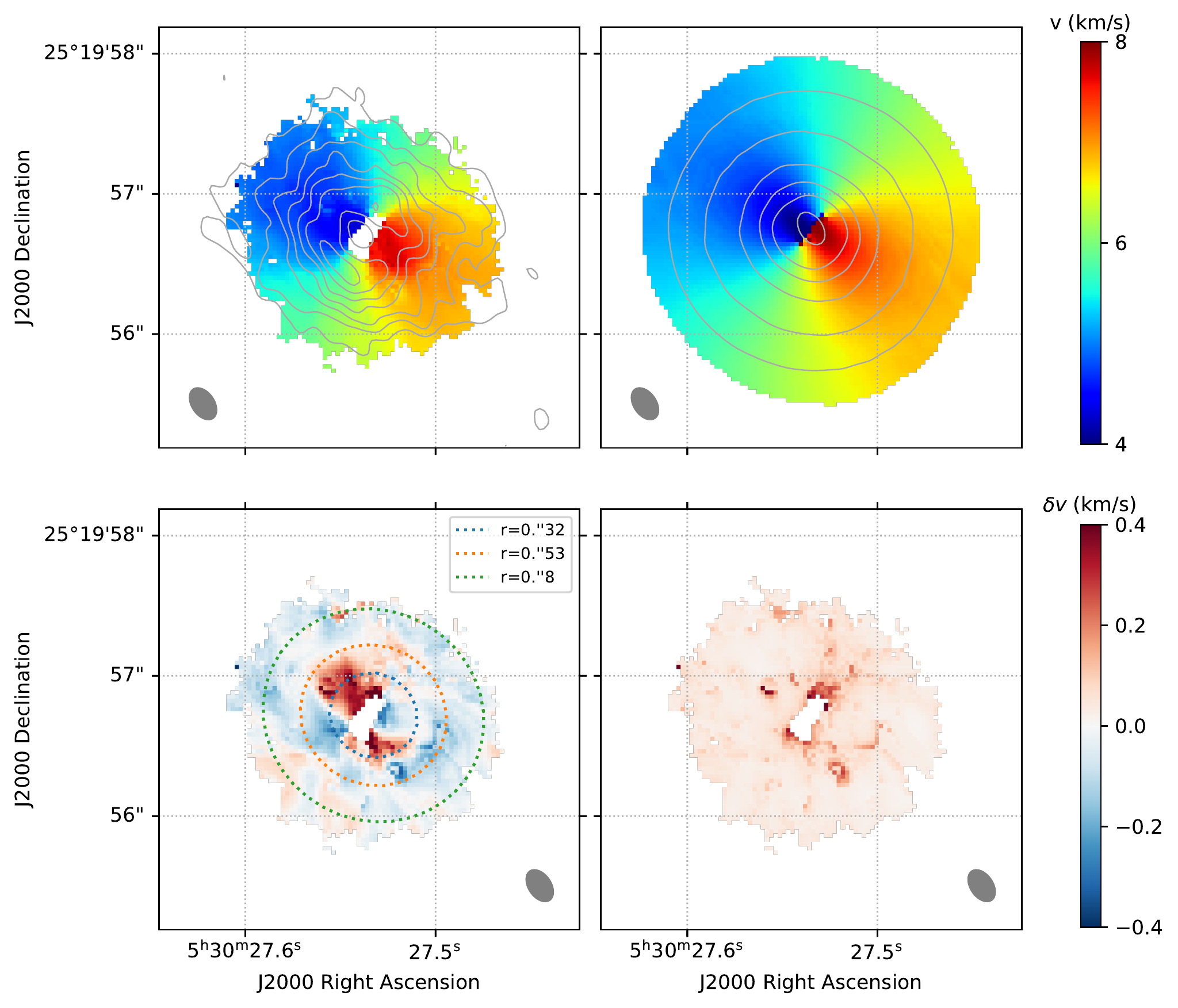}
\caption{$^{13}\mathrm{CO}$ $J=3-2$ integrated intensity (contours) and velocity (color) maps from the ALMA observations (\textit{upper left}) and those of our circular Keplerian disk model (\textit{upper right}) fitted to the observed visibilities using {\tt{DiskFit}}. The model and observed intensity maps have the same beam size, shown as a gray ellipse in the lower corner of each panel. Lower left panel presents the velocity deviations obtained by subtracting the model velocity from the observed velocity maps. Inner to outer dotted ellipses denote the radii of the edge of the inner cavity at 0\farcs32, the outer dust clump at 0\farcs53, and the outer extents of the gas detected in $^{13}$CO $J=3-2$ above 5$\sigma$ at 0\farcs8, respectively. Lower right panel presents the uncertainty in the measured peak velocity. \label{fig:vel_plots}}
\end{figure*}

The line and continuum images generated with the same weighting have been presented and discussed in detail in \citet{Boehler_2018}, so in the present paper, we only present the integrated intensity, peak velocity maps and position-velocity diagram of the $^{13}$CO and C$^{18}$O $J=3-2$ lines (Fig.\ref{fig:vel_plots} and \ref{fig:pv}). 
The peak velocity of each position was measured by fitting a Gaussian function to the observed line profile, and the line emission below half maximum was excluded from the Gaussian fitting. For $^{13}\mathrm{CO}$ $J=3-2$, the Gaussian fitting was performed when the detection is above $5\sigma$ in the velocity channels maps. For $\mathrm{C}^{18}\mathrm{O}$ $J=3-2$, positions with detection above $2\sigma$ were also included in the fitting due to weaker emission.

\begin{deluxetable*}{ccclDlc}[ht!]
\tablecaption{Summary of Images\label{tab:images}}
\tablewidth{0pt}
\tablehead{
\colhead{Line/Continuum} & \colhead{Rest Frequency (GHz)} & \colhead{Beam Size (PA)} & \colhead{Noise (beam$^{-1}$)}}
\startdata
0.88 mm continuum & 343.01 & $0\farcs16\times0\farcs1\ (25.54\arcdeg)$ & 214~$\mu$Jy
\\
$^{13}$CO $J=3-2$ & 330.5879653 & $0\farcs26\times0\farcs17\ (34.18^\circ)$ & 8.7~mJy
\\
C$^{18}$O $J=3-2$ & 329.3305525 & $0\farcs26\times0\farcs17\ (32.95^\circ)$ &  13.1~mJy
\\
\enddata
\tablecomments{Noise is quoted to be per beam per channel for molecular line images.}
\end{deluxetable*}

\par

\section{Analysis and Results}\label{sec:results}

\subsection{Position-Velocity Diagram}\label{sec:pv}

\begin{figure}[ht!]
\epsscale{1.3}
\plotone{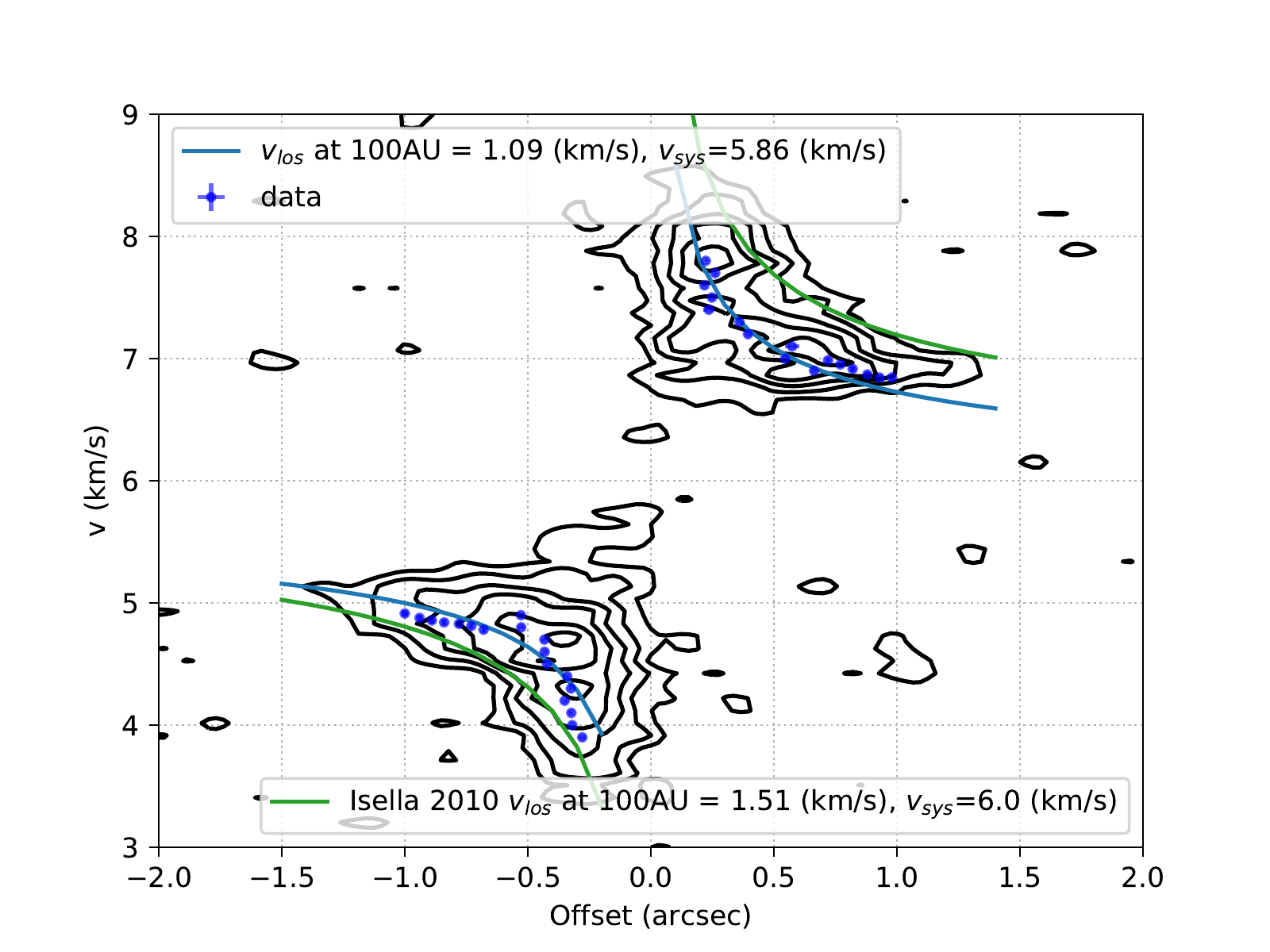}
\caption{Position-velocity diagram of the $^{13}\mathrm{CO}$ emission cut along the major axis of the disk in MWC~758. The contour levels are set from 0.02--0.14 $\mathrm{Jy\ beam}^{-1}$ at intervals of 0.02 $\mathrm{Jy\ beam}^{-1}$. Blue dots denote the observed peak velocities at different positions for lower velocity channels and positions of the intensity peaks at different velocities for higher velocity channels, and blue lines delineate the best-fit rotational curves of a power-law velocity profile with a power-law index of $-$0.5. The projected Keplerian velocity $v_{\rm los}$ at 100 AU is measured to be 1.09~$\mathrm{km\ s}^{-1}$. Green lines delineate the Keplerian velocity profile for the given stellar mass of $2~M_\sun$ and inclination of $21^\circ$ with $v_{sys}=6.0\ \mathrm{km\ s}^{-1}$ from \citet{Isella_2010}.\label{fig:pv}}
\end{figure}

We extracted a position-velocity diagram along the major axis of the MWC~758 disk from the $^{13}\mathrm{CO}$ data cube, as presented in Figure \ref{fig:pv}. We measured the peak velocities at different offsets and the positions of the intensity peaks in different velocity channels to construct a velocity profile. Then we fitted the velocity profile with that of circular Keplerian rotation and measured the projected Keplerian velocity at a radius of 100~AU to be 
$1.09~\mathrm{km~s}^{-1}$ and the systemic velocity to be $5.86~\mathrm{km~s}^{-1}$.
We also calculated the expected line-of-sight velocity at a radius of 100~AU to be $1.51~\mathrm{km~s}^{-1}$ by adopting the inclination angle of $21^\circ$ and stellar mass $2~M_{\sun}$ in the literature \citep{Isella_2010}. There is 28\% difference in our measured and expected velocities, which could hint that the combination of the inclination angle and stellar mass in MWC~758 needs to be reconsidered. 
\par
For a given projected Keplerian velocity, the stellar mass and the inclination angle are degenerate. The inclination of the MWC~758 disk was estimated to be 21$\arcdeg$ from the SED fitting and CO $J=3-2$ observations \citep{Isella_2010}. When the inclination of 21$\arcdeg$ was adopted, we calculated the stellar mass to be 1.0~$M_\sun$ from our measured projected velocity, which is inconsistent with the spectroscopic stellar mass ranging from 1.5 to 1.9 $M_\odot$ \citep{Vioque_2018, Garufi_2018} by 30\% to 45\%. It is also different from the previous estimate of $2.0\pm0.2$~$M_\sun$ from the disk rotation observed in the CO $J=3-2$ line with the Submillimeter Array \citep{Isella_2010}. We note that the $^{13}$CO line often traces an upper layer with a scale height of 0.1--0.2 in a protoplanetary disk \citep[e.g.,][]{Pinte_2018}, and the Keplerian velocity in the midplane could be underestimated. Nevertheless, we found that this correction for the possible scale height cannot reconcile this discrepancy.

\subsection{Circular Keplerian Disk Model}\label{subsec:circmodel}
To measure the gas kinematics in the MWC~758 disk and search for deviation from circular Keplerian rotation, we fitted the $^{13}\rm{CO}$ and $\rm{C}^{18}\rm{O}$ data with circular Keplerian disk models using the software package {\tt{DiskFit}} \citep{DiskFit}. The fitting was performed separately to the two molecular lines. With {\tt{DiskFit}}, radiative transfer calculations were conducted, and the Monte Carlo Markov Chain method was applied to search for the best fits to the observed visibility data. The column density, temperature, and scale height in the disk models are characterized by power-law functions of radius. The radial profile of the rotational velocity ($v_{rot}$) is fixed to be Keplerian rotation, meaning that $v_{rot}\propto r^{-0.5}$. The free parameters in our model fitting include position angle of the major axis (PA), inclination, systemic velocity, stellar mass, and radial profiles of density and temperature of the molecular gas, as listed in Table \ref{tab:param}. 
Details of the radiative transfer calculations, disk parameterization, and minimization methods are described in \citet{DiskFit}. Finally, we generated model images from the best-fit visibility data, and we cleaned, smoothed, and regridded the model images to match the observed images.

\par

From test runs of the fitting, we found that the inclination and the stellar mass cannot be simultaneously constrained with the current data. Since we are not able to disentangle these two parameters, which are degenerate in determining the projected Keplerian velocity pattern, in Table \ref{tab:param} we present the parameter results of two sets of the best-fit models, one with the inclination fixed at the literature value of $21^\circ$ and one with the stellar mass fixed to be $2~M_\odot$ \citep{Isella_2010}. We confirmed that the projected velocity maps from these two models are indeed consistent within the uncertainty in the regions of our interest (Appendix~\ref{deg_i-Ms}).
The integrated intensity and velocity maps of the best-fit models of the $^{13}$CO data with the fixed inclination of $21^\circ$ are presented in the upper-right panel of Figure \ref{fig:vel_plots}. The fitting results of the $^{13}$CO and C$^{18}$O emission lines agree within the uncertainties.  The C$^{18}$O results have larger uncertainties due to the lower S/N of the C$^{18}$O data, so below we focus on the $^{13}\mathrm{CO}$ results. The C$^{18}$O results are shown in Appendix \ref{c18o_fitting}. 

\par

We have also performed fitting using the software package {\tt{Eddy}} \citep{eddy}, which fits the observed velocity map with analytical models without further considering the uv coverage and intensity distribution. We found that {\tt{Eddy}} also could not fit the stellar mass and inclination of MWC~758 simultaneously, and the solution was not converged, the same as our fitting using {\tt{Diskfit}}. If we fixed the stellar mass to be $2~M_\sun$ or the inclination angle to be $21^\circ$, the fitted inclination angle and stellar mass from {\tt{Eddy}} are consistent with our fitting results using {\tt{Diskfit}}, but with larger error bars.
Considering the limited angular resolution of our data compared to the size of the eccentric dust ring, we adopted the fitting results and model images from {\tt{Diskfit}} for our subsequent analysis to avoid any possible bias due to the beam convolution.
Molecular-line data with better sensitivity and resolutions are needed to measure the inclination and stellar mass of MWC~758 without any degeneracy from the disk rotation. 

\par

The negative sign of the inclination in our results (Table \ref{tab:param}) means that the northwestern part of the disk is the near side, which was also suggested by \citet{Boehler_2018}. With the fixed inclination of $21^\circ$, the best-fit stellar mass is $1.21\pm0.01~M_{\sun}$, which is comparable to that estimated with our analytical analysis of the velocity structures in the position-velocity diagram in Section \ref{sec:pv}.  Nevertheless, this best-fit stellar mass is lower than $2.0\pm0.2~M_{\odot}$ estimated\footnote{To further investigate this discrepancy with the stellar mass and inclination ($2~M_\odot$ and $i=21^\circ$) estimated in \citet{Isella_2010} from the SMA $^{12}\mathrm{CO}\ \mathrm{J=3-2}$ data, we additionally made three model images of the $^{12}\mathrm{CO}\ \mathrm{J=3-2}$ emission with different combinations of the stellar mass and inclination, (1) $2~M_\odot$ and $i=16^\circ$, (2) $1.2~M_\odot$ and $i=21^\circ$, and (3) $2~M_\odot$ and $i=21^\circ$, using {\tt{Diskfit}}. We convolved the model images to the resolution of the SMA observations and generated velocity maps. We compared these velocity maps by subtracting one from the other. We found that with the given resolution and sensitivity of the SMA observations, it is difficult to detect the velocity differences among the three models clearly. Thus, the uncertainty in the previous estimate may be underestimated.} by \citet{Isella_2010} and also lower than the spectroscopic mass, as discussed above. Our best-fit systemic velocity of $5.93\pm0.01$~km~s$^{-1}$ is consistent with that estimated by \citet{Boehler_2018}, $5.90\pm0.05$~km~s$^{-1}$. We note that in our best-fit disk model, the gas density has a steep radial profile with a power-law index of 2.4, which is slightly larger than typical values between 0 and 2 \citep{PPD_annrev}.  With only one transition of the $^{13}\mathrm{CO}$ and $\mathrm{C}^{18}\mathrm{O}$ lines we cannot unambiguously constrain the density and temperature profiles at the same time. The slope of the density profile is coupled with the column density and inner- and outermost radii in our model fitting.  Thus, our best-fit density and temperature profiles should be considered as a possible solution that can approximately match the observed intensity distribution, and our main goal is to reproduce the Keplerian rotation in the disk but not to infer the physical conditions.
\par
 The other solution with the fixed stellar mass of $2~M_\sun$ gives a systemic velocity consistent with the previous solution within the error bars, but the inclination in this case is $-16^\circ$. As the slope of the column density and the inner and outer disk radii are also coupled parameters, this solution returns a larger inner radius that better reproduces the size of the inner cavity in MWC~758, leading to a steeper column density profile with a power-law index of 6.4. Nevertheless, our model does not account for the presence of the clumps, spirals, and rings in the MWC~758 disk. The density structures in the MWC~758 disk have been discussed in more detail in \citet{Boehler_2018}. In the present paper, we focus on the gas kinematics in the MWC~758 disk. 

\par
In order to compare with the observational results of the eccentric dust ring analyzed by adopting an inclination of $-21\arcdeg$ in \citet{Dong_2018}, the solution with the inclination of $-21\arcdeg$ and $M_\star$ of 1.2~M$_\sun$ is adopted in our further discussions of the gas kinematics. As discussed below, the results of our subsequent analysis are not affected by the adopted inclination angle, and the difference in the projected velocity maps of the two best-fit models is comparable to the observational uncertainty at radii larger than $0\farcs3$, which is the region of our interest (Appendix~\ref{deg_i-Ms}). We measured the eccentricity of the dust ring deprojected by an inclination of $-16\arcdeg$ to be 0.1 with a semi-major axis of $0\farcs32$, consistent with the results in \citet{Dong_2018}. The effect of the different inclination on the dust ring eccentricity is unresolvable and does not impact our discussion.

\begin{deluxetable}{cchlDlccc}[ht!]
\tablecaption{Summary of Best-Fit Model\label{tab:param}}
\tablewidth{0pt}
\tablehead{
\colhead{Parameter (units)} & \colhead{Fixed Inclination} & \nocolhead{} & \colhead{Fixed Stellar Mass} }
\startdata
PA ($\arcdeg$) & $60.6\pm0.3$ &  & $59.8\pm0.3$
\\
inclination ($\arcdeg$) & $-21.00$ &  & $-16.3\pm0.1$
\\
$v_{\rm sys}$ (km~s$^{-1}$) & $5.94\pm0.01$ &  & $5.92\pm0.01$
\\
$M_{\star}$ ($M_{\odot}$) & $1.20\pm0.01$ &  & $2.00$
\\
$r_{in}$ (au) & $17.8\pm2.3$ &  & $41.6\pm2.4$
\\
$r_{out}$ (au) & $168.0\pm1.3$ &  & $177.4\pm2.5$
\\
$\log\Sigma_0$ (cm$^{-3}$)& $17.1\pm0.3$ &  & $17.2\pm0.2$
\\
$pp$  & $2.4\pm0.6$ &  & $6.4\pm1.0$
\\
$T_0$ (K) & $25.7\pm0.4$ &  & $24.3\pm0.5$
\\
$qq$ & $0.55\pm0.07$ &  & $0.76\pm0.09$
\\
$H_0$ (au) & $16.1\pm3.0$ &  & $20.1\pm2.9$
\\
$hh$ & $-1.3\pm1.4$ &  & $-1.9\pm1.6$
\enddata
\tablecomments{PA is the position angle of the major axis of the disk. Inclination is defined as the angle between the normal axis of the disk and the line of sight. $v_{\rm sys}$ is the systemic velocity. Power-law profiles of the disk properties are parameterized as: (1) column density $\Sigma(r)=\Sigma_0(\frac{r}{r_0})^{-pp}$, (2) temperature $T(r)=T_0(\frac{r}{r_0})^{-qq}$, and (3) scale height $H(r)=H_0(\frac{r}{r_0})^{-hh}$. These three parameters are optimized to best reproduce the observed intensity distribution. We note that these fitted parameters may not reflect the actual density, temperature, and height profiles in the MWC 758 disk because the actual profiles could be be more complex than the simple power-law functions adopted in the model \citep{Pinte_2018T, Calahan_2021, Law_2021}, and the density and temperature cannot be unambiguously constrained at the same time with only one transition. 
} The reference radii $r_0$ are all taken to be 100 AU. $r_{in}$ and $r_{out}$ denote the inner edge and outer extents of the gas.
\end{deluxetable}


 Figure \ref{fig:vel_plots} shows the observed velocity map (upper left), the Keplerian model velocity map (upper right), the residuals after subtracting the model velocity maps from the observed velocity maps (lower left), and the uncertainty in the velocity (lower right) of the $^{13}$CO $J=3-2$ emission. The ellipses denote the reference radii for the following discussion. The innermost ellipses delineate the radius of the eccentric dust ring at the edge of the inner cavity, 0\farcs32. The middle ellipses encircle the position of the dust clump at a radius of 0\farcs53 at the northwestern side of the disk. The outermost ellipses show the outer extents of the disk with $\ge5\sigma$ detection in $^{13}$CO $J=3-2$, at a radius of 0\farcs8. 
 
In the residual velocity map of $^{13}$CO, we can see clear redshifted excesses at radii $< 0\farcs53$, especially along the major and minor axes. The deviations are 2 to 4 times larger than the velocity channel width of 0.1 km~s$^{-1}$, and around 10\% of the local Keplerian velocity. The uncertainties in the fitted peak velocity due to the noise are also more or less an order of magnitude smaller than the velocity deviations. In outer parts of the disk, the velocity deviations are only $\sim\pm0.1$~km~s$^{-1}$, comparable to the velocity channel width and therefore negligible. The C$^{18}$O emission exhibits a similar trend at radii $< 0\farcs53$, although the significance of the velocity deviations in $\mathrm{C}^{18}\mathrm{O}$ is relatively low ($<$2$\sigma$) due to its lower signal-to-noise ratio (Appendix~\ref{c18o_fitting}).

\begin{figure*}[ht!]
\epsscale{1.2}
\plotone{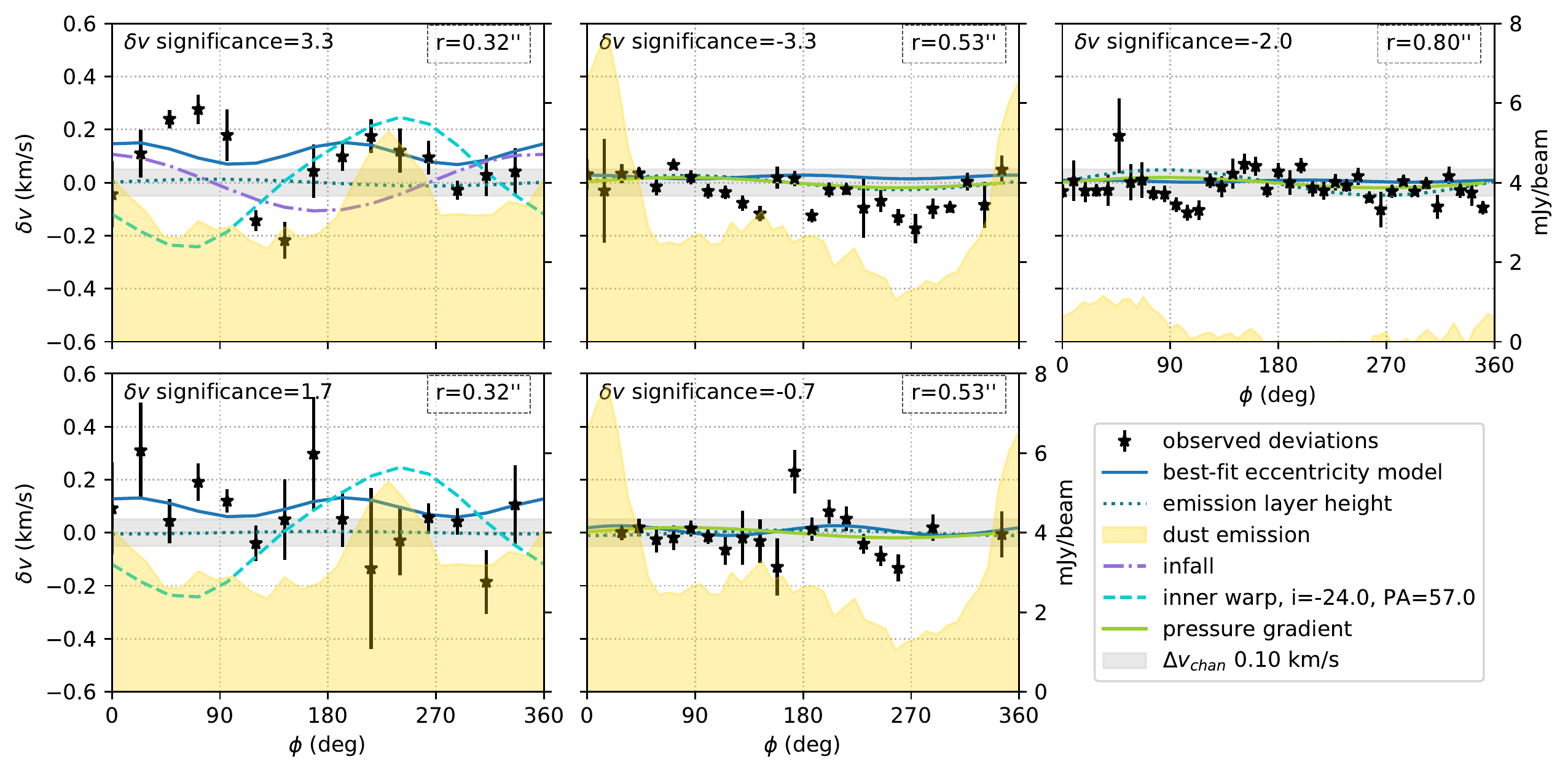}
\caption{Azimuthal profiles of the velocity deviations from the Keplerian rotation at a radius of 0\farcs32 (\textit{left panels}), 0\farcs53 (\textit{middle panels}), and 0\farcs8 (\textit{right panel}) measured from the $^{13}\mathrm{CO}$ ${J=3-2}$ (\textit{upper panels}) and $\mathrm{C}^{18}\mathrm{O}$ ${J=3-2}$ (\textit{lower panels}) data. $\phi=0^\circ$ refers to the northwestern side of the minor axis.  Due to the limited S/N of the $\mathrm{C}^{18}\mathrm{O}$ emission, the deviations at $r=0\farcs8$ for this line are not shown. Overlays are the profiles of the expected velocity deviations of different mechanisms causing non-Keplerian motions. The considered mechanisms include the height of the emitting layer, radial flows, an inner warped disk, pressure gradient, and an eccentric orbit. The mean S/N ratio of the observed velocity deviations is shown in the upper left corner of each panel. The gray shading denotes the velocity resolution of the observations. The yellow shaded background shows the 0.88 mm continuum intensity. \label{fig:azimuth_vel}}
\end{figure*}

\subsection{Origin of Velocity Deviations}\label{sec:dev_origins}
We extract the azimuthal profiles of the velocity deviations as shown in Figure \ref{fig:azimuth_vel} to discuss various possible mechanisms which may cause the non-Keplerian rotation seen in Figure \ref{fig:vel_plots}, including pressure gradient, height of the emitting layer, radial flow, a warped inner disk, or eccentricity. $\phi=0^\circ$ in Figure \ref{fig:azimuth_vel} refers to the northwestern side of the minor axis. The separation between the data points is half the beam width. Due to the larger error bars of the C$^{18}$O velocity, the velocity deviations observed in C$^{18}$O have low significance, all less than 2$\sigma$, and will not be included in the following discussion. For $^{13}$CO, the local velocity exhibits significant deviations from the circular Keplerian rotation along the inner dust ring at a radius of $0\farcs32$, with a mean S/N of $3.3\sigma$. Deviations along the major axis of the disk and neighboring regions are positive, while some blue-shifted residuals can be seen around the minor axis. At a radius of 0\farcs53, most of the velocity deviations are less than 0.2~km~s$^{-1}$ and negative. Velocity deviations at larger radii are comparable with the velocity channel width, denoted within the gray shading in Figure \ref{fig:azimuth_vel}. 

\subsubsection{Pressure Gradient}
When the pressure gradient in a disk is directed outward, the gas motion in the disk becomes sub-Keplerian, and vice-versa if the pressure gradient is directed inward. 
The observed deviations at $r=0\farcs53$ follow a trend similar to the expected deviation due to the pressure gradient, which is a sinusoidal curve with positive deviations around $\phi=90\arcdeg$ and negative deviations around $\phi=270\arcdeg$. 
We adopted the gas density and temperature profiles from our best-fit model to estimate expected deviations due to the pressure gradient at outer radii.
We found that although the trend is similar, the magnitude of the observed deviation at $r=0\farcs53$ is larger than the expectation (Figure~\ref{fig:azimuth_vel} middle panel). Nevertheless, there could be additional variation in the pressure not included in our model. For example, the clump seen in the continuum image at $r = 0\farcs43$ and a position angle of $270\arcdeg$ could suggest dust trapping in a local pressure maximum \citep{Casassus_2019}. Thus the local pressure maximum could cause a larger velocity deviation than that in our calculations with the axisymmetric disk model. 

On the contrary, the observations of the inner cavity and innermost dust ring in the MWC~758 disk, suggest the presence of a pressure maximum at a radius of 0\farcs32, which could halt the radial drift of dust at the outer edge of the cavity \citep{Dong_2018}. Therefore, it can be inferred that the velocity deviations at $r=0\farcs32$ are unlikely due to pressure gradient.

\subsubsection{Emission Height}

Since the assumed density and temperature profiles in our Keplerian disk model are simple, the fitted scale height may not reflect the actual scale height from the hydrodynamic equilibrium in the disk. Thus, to further test if any velocity residual is due to inaccurate height of the emitting layer, we also use an alternative method, as introduced in \citet{Pinte_2018}, to estimate the height of the emitting layer. We measured intensity weighted mean positions on the two sides of the disk with respect to the major axis, and only the pixels above $5\sigma$ detection were included in the calculations. We fitted the heights with a power-law profile
with a power-law index of 2 \citep{PPD_annrev}, and the height of the emitting layer is estimated to be 20 AU at a radius of 100 AU for $^{13}\mathrm{CO}$.

The effect of the height of the $^{13}\mathrm{CO}$ emitting layer is then considered by modifying the Keplerian speed as,  
\begin{equation}
    v_{\rm kep,z} = \sqrt{\frac{GM_{\star}}{r^2+z^2}\frac{r^2}{(r^2+z^2)^{1/2}}},
    \label{eq:vkeph}
\end{equation}
where $G$ is the gravitational constant, $z$ is the height of the emitting layer, and $v_{\rm kep,z}$ is the Keplerian velocity at a radius of $r$ and a height of $z$.
The rotational velocity of gas in a layer above the midplane is then projected onto the line of sight, and subtracted from the best-fit midplane Keplerian rotation. The corresponding deviations are shown in Figure \ref{fig:azimuth_vel}.

\par
The possible velocity deviations due to the height of the emitting layer are estimated to be $\delta v= 0.04$~km~s$^{-1}$ at $r=0\farcs32$ and at most $\delta v=0.09$~km~s$^{-1}$ at $r=0\farcs8$ along the major axis, projected along the line of sight, which are minimal compared to the observed deviations.

\subsubsection{Inner Warped Disk}
Previous millimeter molecular-line and near-infrared studies have found that the inner disk is warped \citep{Eisner_2004, Isella_2006, Isella_2008, Boehler_2018}. 
The high-resolution ALMA continuum observations have directly imaged the inner disk and measured the inclination to be $50\arcdeg\pm6\arcdeg$ and PA to be $7\arcdeg\pm9\arcdeg$ \citep{Francis_2020}. Twisted iso-velocity curves have also been observed in the $^{13}$CO emission within a radius of 0\farcs1 in the MWC~758 disk, indicative of a misaligned inner disk \citep{Boehler_2018}.

\par
If the disk gradually becomes warped from the outer to inner radii, the disk rotation at $r=0\farcs32$ could have different inclination and PA from those in the outer disk, such changes in the disk orientation could introduce velocity deviations. We computed the expected line-of-sight velocity of a Keplerian rotation assuming different inclination and PA along the reference ellipse and subtracted it from that of our best-fit Keplerian rotation with the inclination of $21\arcdeg$ and PA of $61\arcdeg$. 
We assumed that the inclination and PA at a radius of  0\farcs32 are between those of the inner and outer disks. 
We tried several different combinations of inclination and PA in these ranges and compared the expected velocity deviation due to the possible warp with the observations. In Figure \ref{fig:azimuth_vel}, we show an example of the velocity deviation if the disk rotation at $r=0\farcs32$ actually has inclination of $24\arcdeg$ and PA of $57\arcdeg$.  
The azimuthal profile of the velocity deviation due to the warp is also a sinusoidal curve, where the deviation is positive on one side of the disk and negative on the other side.
We could not find any combinations of inclination and PA to explain the overall trend of the observed velocity deviations at $r=0\farcs32$, which are mostly positive. 

\subsubsection{Radial Flows}
The twist in the iso-velocity contours of the $^{13}$CO and $^{12}$CO emission seen in the inner 0\farcs1 region in the MWC~758 disk observed with ALMA \citep{Boehler_2018, Isella_2010}, could also suggest inflow \citep{Rosenfeld_2014}. We computed expected velocity deviations due to radial motion, assuming an axisymmetric infalling motion \citep{Rosenfeld_2014}. The radial flows can maximally reach free-fall velocity.
In our calculations, we assumed the radial velocity to be 10\% of the Keplerian velocity. The results are shown in Figure \ref{fig:azimuth_vel}. 
\par
As shown in Figure \ref{fig:azimuth_vel}, the global feature of the observed velocity deviations does not agree with that caused by the axisymmetric radial motion. The velocity deviation due to the radial motion follows a sinusoidal curve with maximum line-of-sight deviations along the minor axis, but in our observation, the maximum deviations occur around $90\arcdeg$ and $225\arcdeg$. It is possible that the radial motion is not axisymmetric. We note that the observed velocity deviation at an azimuth angle of $\sim135\arcdeg$ is substantially more negative than that at other azimuth angles. The presence of an infalling flow at azimuth angles around $135\arcdeg$ may explain the observed negative velocity deviation at those azimuthal angles.

\subsubsection{Eccentric Disk}\label{subsec:ecc}

\begin{figure*}[ht!]
\epsscale{1.2}
\plotone{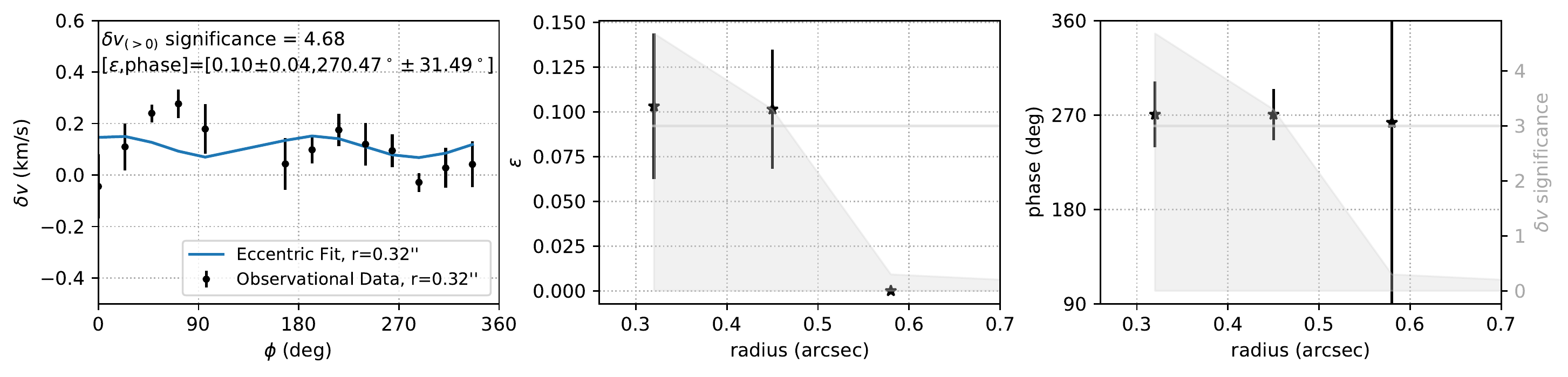}
\caption{(\textit{Left}) Our fitting of eccentric gas motion to the observed velocity deviations in the ALMA archival data of the $^{13}$CO emission in the MWC~758 disk at $r=0.\!\!^{\prime\prime}$32, where the inner dust ring has been found to be eccentric. Only the positive observed deviations are considered in the fitting because the expected deviations from an eccentric orbit are positive. $\phi=0^\circ$ refers to the northwestern side of the minor axis. (\textit{Center}) The radial profile of our derived gas eccentricity from the archival $^{13}\mathrm{CO}$ data. Data points are separated by half the beam size. (\textit{Right}) The radial profile of the best-fit position angle of the pericenter of the eccentric orbit from the archival $^{13}\mathrm{CO}$ data. In the central and right panels, the gray shaded area shows the signal-to-noise ratio of the measured velocity deviations, and the gray line denotes the $3\sigma$ confidence level of the velocity deviation. \label{fig:13COecc}}
\end{figure*}

The outer edge of the inner cavity in the MWC~758 disk is eccentric with an eccentricity of 0.1, as observed in the 0.88 mm continuum emission \citep{Dong_2018}. Therefore, if the gas is also dynamically perturbed into an eccentric orbit, the eccentricity may cause the observed velocity deviations. To compute the velocity deviations from circular Keplerian rotation due to a small eccentric gas motion, we applied the eccentric model in Equation \ref{eq:ecc} \citep[e.g.][]{ Hsieh_2012, Goodchild_Ogilvie_2006},

\begin{equation}
\begin{cases}
    \delta v_{r} = e_g v_{kep} \sin\omega\\
    \delta v_\phi = \frac{1}{2}e_g v_{kep}\cos\omega,
\end{cases}
\label{eq:ecc}
\end{equation}

\noindent where $\delta v_r$ and $\delta v_\phi$ are the radial and azimuthal velocities deviating from circular Keplerian rotation, $e_g$ is the amplitude of gas eccentricity, and $\omega$ is the phase angle, starting from 0 at the pericenter of the eccentric orbit. 
The eccentric rotational velocity is faster at the pericenter and slower at the apocenter than the circular Keplerian velocity. As a result, if the pericenter of an eccentric orbit is at the southwestern side of the MWC~758 disk, where the Keplerian velocity is redshifted along our line of sight, the projected velocity deviations due to the eccentric gas orbit would all be positive. In contrast, if the pericenter of an eccentric orbit is at the northeastern side of the MWC~758 disk, the projected velocity deviations would all be negative.
Thus, the presence of eccentric gas orbital motions with the pericenter at the southwestern side of the MWC~758 disk may explain the observed velocity deviations at a radius of 0\farcs32, which are mostly positive.

\par

On the assumption that the observed positive deviations at $r = 0\farcs32$ are due to eccentricity, we performed fitting of the eccentric model (Eq.~\ref{eq:ecc}) to the observed deviations that are positive within the error bars to estimate the gas eccentricity $e_g$ and the position angle of the pericenter in the MWC~758 disk. Since the S/N of the C$^{18}$O emission is too low for a robust constraint, we have only performed the fitting on the $^{13}$CO data and the results are shown in Figure \ref{fig:13COecc}.
We found the best-fit eccentricity at $r=0\farcs32$ to be  $0.1\pm0.04$ and the position angle of the pericenter to be  $270\arcdeg\pm32\arcdeg$.
We have also tested and found that including the negative deviations in our fitting with the eccentric model does not change the best fit of the position angle of the pericenter and the eccentricity at $r=0\farcs32$. We also performed the same fitting to the positive velocity deviations at outer radii (Figure \ref{fig:13COecc}), but there is no detectable eccentricity or any significant velocity deviation at radii larger than 0\farcs45.

\section{Discussion}\label{sec:discuss}

\subsection{Non-Keplerian Motion}
\par
It can be seen from the previous analysis that the expected velocity deviations from the Keplerian rotation due to the pressure gradient and the height of the emitting layer are negligible compared to the observed deviations, and the warp, inflow, or their combinations, which have sinusoidal profiles with both positive and negative velocity deviations, also cannot explain the observations showing the positive velocity deviations at most of azimuthal angles at $r = 0\farcs32$.
The velocity deviations due to an eccentric gas motion with its percenter at the soutwestern side of the MWC~758 disk are all positive.
Thus, the overall trend of the observed velocity deviations at $r = 0\farcs32$ is best explained with the eccentric gas motion, especially the presence of positive deviations at azimuthal angles both around 45$\arcdeg$--90$\arcdeg$ and 180$\arcdeg$--270$\arcdeg$,
and the inflow may be able to account for the deviation at an azimuth angle of $\sim135\arcdeg$ (Fig.~\ref{fig:azimuth_vel}). The observed velocity deviations could be due to a superposition of these two mechanisms.
\par
Assuming that the positive velocity deviations at $r = 0\farcs32$ are due to the eccentric gas motion, 
 our best-fit gas eccentricity and position angle of the pericenter of $e_g=0.1\pm0.04$ and $270\arcdeg\pm32\arcdeg$, respectively, are consistent with those of the eccentric inner dust ring with the eccentricity $e_d$ of $0.1\pm0.01$ and the pericenter at the position angle of $275\arcdeg\pm10\arcdeg$ \citep{Dong_2018}.
 In addition, we tentatively observe a decrease of the eccentricity going further out from the cavity in the MWC~758 disk. 
\par
 
 We note that although our eccentric model cannot fully explain the observed velocity deviations and even the positive deviations could be contaminated by other mechanisms, our estimate of the eccentricity of the MWC~758 disk can be still considered as an upper limit. If the eccentricity is actually larger, we expect to detect larger velocity deviations, but the observed velocity deviation is at most 10\% of the Keplerian velocity. We also note that the inclination of the MWC~758 disk is not well constrained as discussed in Sec. \ref{sec:dev_origins}. 
 Nevertheless, our estimate of the eccentricity is independent of the adopted inclination angle. This is because the eccentricity is constrained from the ratio of the velocity deviation to the Keplerian velocity, and thus the projection effect is cancelled out. 
 We have tested it by performing the same analysis with the best-fit Keplerian disk model with the inclination of 16$\arcdeg$ and $M_\star$ of 2~$M_\sun$, and the estimated eccentricity is indeed consistent with that estimated with the inclination of 21$\arcdeg$ and $M_\star$ of 1.2~$M_\sun$ within the uncertainty (Appendix~\ref{ecc_i16}). Furthermore, the pattern of the velocity deviations due to inaccurate stellar mass or inclination is different from that due to an eccentric motion \citep{Yen_2020}, and we have also confirmed that the velocity difference between our two best-fit models with the different inclination and stellar mass do not lead to a non zero eccentricity in our analysis (Appendix~\ref{deg_i-Ms}).

\begin{figure*}[ht!]
\epsscale{1.15}
\plotone{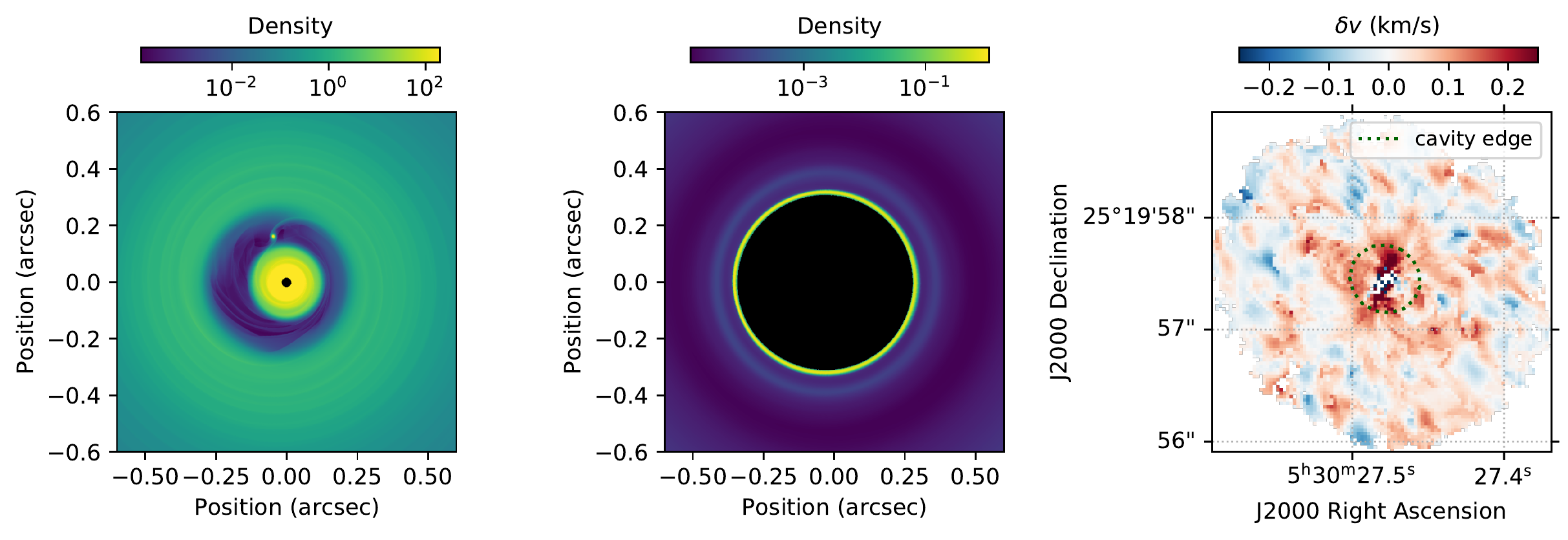}
\caption{(\textit{Left}) Face-on view of the gas density distribution in our hydrodynamic simulation. An eccentric gas cavity is seen. Spatial axes are in units of arcsecond, and the same distance to MWC~758 is adopted. (\textit{Middle}) Face-on view of the dust density distribution in our hydrodynamic simulation. A thin ring of dust accumulation is seen at the cavity edge. (\textit{Right}) Velocity deviations from the Keplerian rotation measured from the velocity map of our synthetic data of the hydrodynamic simulation. An open ellipse delineates the edge of the inner cavity. The velocity deviations are redshifted at all azimuthal angles around the edge of the inner cavity, which is theoretically expected. \label{fig:ecc_simobs_map}}
\end{figure*}

\begin{figure*}[ht!]
\epsscale{1.15}
\plotone{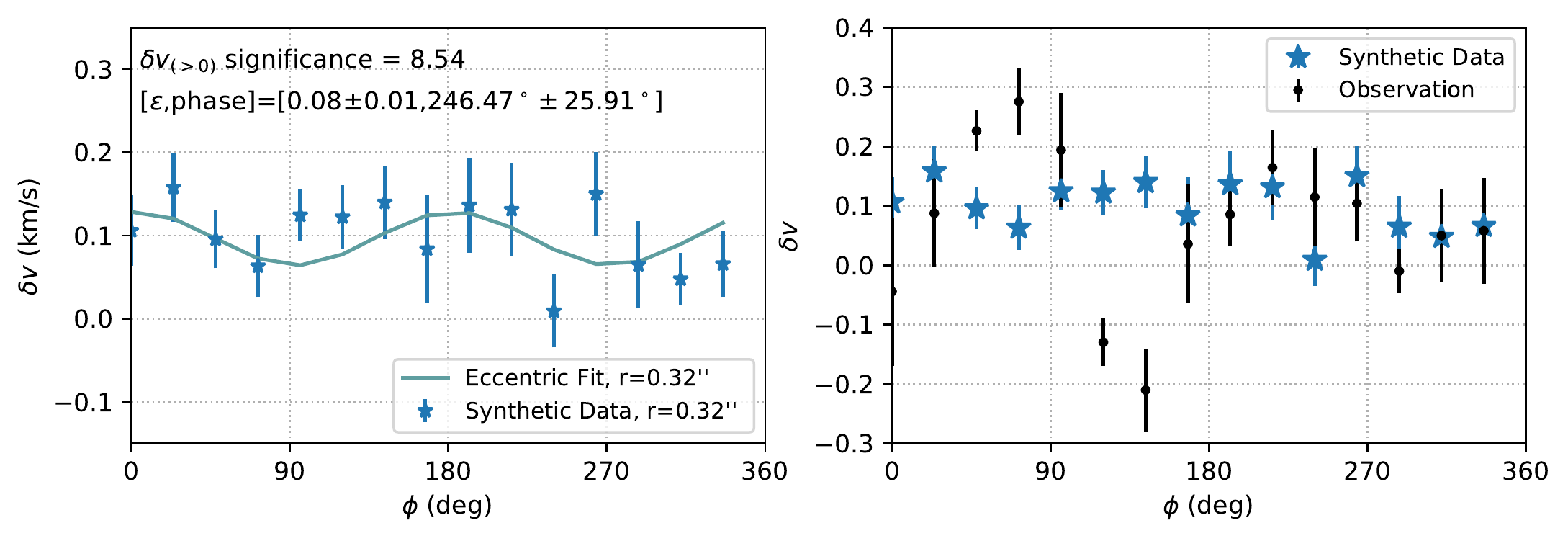}
\caption{
(\textit{Left}) Our fitting of the eccentric gas motion (light blue curve) to the measured velocity deviations in the synthetic data (blue data points). (\textit{Right}) Comparison of the velocity deviations around the inner cavity between the observed and synthetic data, which are shown as black and blue data points, respectively. $\phi=0^\circ$ refers to the northwestern side of the minor axis of the disk.\label{fig:ecc_simobs_vel}}
\end{figure*}


\par
To further demonstrate the expected velocity deviations due to the eccentric gas motion in a protoplanetary disk, 
we conducted hydrodynamic simulations of a protoplanetary disk with an embedded gas giant planet using the FARGO3D code \citep{FARGO3D}.
The simulation was carried out in 2D with the two-fluid approximation \citep{2fluid} to simulate both dust and gas. 
In our simulation, the range of the radial domain was adopted to 0.1 $r/r_p$ to 10 $r/r_p$ with a number of 1500 zones in log space, where $r_p$ is the distance between the planet and the star, while the azimuthal domain was equally splitted into 2048 zones. 
In the MWC~758 disk, the brightness temperatures of the 0.88 mm continuum and $^{13}$CO (3--2) emission were measured to be 3.0 and 25.8 K at a radius of $0\farcs5$, respectively. The 0.88 mm continuum emission most likely traces the disk midplane and may be optically thin \citep{Boehler_2018}, and the $^{13}$CO (3--2) emission is expected to be optically thick and trace the disk surface. 
Thus, their brightness temperatures can be adopted as the lower and upper limits of the temperature in the disk midplane, 
and the corresponding pressure scale heights ($h/r$) in the hydrodynamical equilibrium, ${h}/{r}={c_s}/{v_{\mathrm{Kep}}}$, were estimated to be 0.03 and 0.08, where $c_s$ is the thermal sound speed and $v_{\mathrm{Kep}}$ is the Keplerian velocity. 
The estimated $h/r$ in the MWC~758 disk is consistent with the expectation from the hydrodynamical models where disk eccentricity is able to develop \citep{Muto_2011, DSHARP_VII}.
Thus, in our simulations, the aspect ratio of the disk was set as $\frac{h}{r}=0.05\times (\frac{r}{r_p})^{0.25}$, an arbitrary value within the estimated range. We note that the aspect ratio here differs from the fitting parameter $H_0$ in Table \ref{tab:param}, which was optimized to match the height of the molecular-line emitting layer. The slope of the initial gas density profile was adopted to be $-3$, similar to our best-fit disk model of MWC~758, and the self gravity of the disk is ignored in the simulation. 
The planet was fixed on a circular orbit with a mass of 0.3\% of $M_\star$, and the accretion of the planet was not considered. The $\alpha$ viscosity was adopted to be $10^{-4}$, and dust feedback was neglected. As for the dust particles, the grain size was fixed in the simulations. The initial radial profile of the Stokes number was 0.1 at $r_p$, where the Stokes number is defined as the ratio of the stopping time to the orbital time for a dust grain experiencing aerodynamic drag \citep[e.g.][]{Weidenschilling_1977, Testi_2014},
and the profile was allowed to evolve in the simulations. 
\par
We extracted the simulation results when the planet rotated for 4300 orbits, 
and an inner cavity and a dust ring formed at 1.9 $r_p$. 
The dust ring is eccentric with an eccentricity of 0.1, the same as that in the MWC~758 disk.
The gas and dust distributions in the simulation are shown in the left and middle panels in Figure~\ref{fig:ecc_simobs_map}.
The radial domain of our simulation is much larger than the size of the inner cavity and dust ring, so the results in the regions of our interest are not affected by the boundary of the computation domain. 
We have also confirmed that the results from simulations with even higher resolutions do not show any significant differences. 
We note that the disk structures in our hydrodynamical simulation are simple and different from the observations of MWC~758 showing clumps and spirals \citep[e.g.,][]{Baruteau_2019, Calcino_2020, Ren_2020}. The goal of our hydrodynamical simulation is to demonstrate velocity deviations due to eccentric motions and test if the deviations can be detected with our data and analysis, and we do not aim to reproduce all the observed features. As shown in the literature \citep[e.g.,][]{Baruteau_2019, Calcino_2020, Ren_2020}, more sophisticated hydrodynamical models are needed to reproduce the complex structures in the MWC~758 disk.
\par

Finally, we scaled our simulation results to have $M_\star$ of 1.2~$M_\sun$ and the radius of the dust ring of 0\farcs32, the same as MWC~758, and the simulation results were also rotated to have a pericenter of the eccentric dust ring at the position angle of 270$\arcdeg$ after projection.
Then we computed the radiative transfer equation with the geometrically thin slab approximation and the assumption of the local thermal equilibrium to generate model $^{13}$CO images, and the same inclination and PA as those in MWC~758 were adopted.
Because our hydrodynamic simulation was conducted with the vertically isothermal assumption, 
we linearly scaled up the temperature profile, as $^{13}$CO often traces an upper warmer layer, so that the peak intensity of the resultant model images matched that in the observations. 
Then we simulated the ALMA observations on our model images with the CASA simulator. 
The right panel in Figure \ref{fig:ecc_simobs_map} presents the results of our imaging simulations, and our synthetic images have approximately the same S/N and resolution as the observations. 
In our synthetic data, the velocity residuals are all positive with a maximum deviation of $\sim0.2~\mathrm{km\ s}^{-1}$ around the inner cavity, similar to the trend in the observations, although the maximum deviation is 50\% lower than that in the observations of $\sim0.3~\mathrm{km\ s}^{-1}$.
\par
We extracted the azimuthal profile of the velocity deviation from the Keplerian rotation from the synthetic data and analyzed it with the same methods presented in Section \ref{subsec:ecc}.
The eccentricity and the position angle of the pericenter were measured to be $0.08\pm0.01$ and $246^\circ\pm26^\circ$ from the synthetic data, respectively. 
These values are consistent with those in our hydrodynamic simulations within the 1$\sigma$--2$\sigma$ uncertainties.
In Figure \ref{fig:ecc_simobs_vel}, we also compare the velocity deviations in the observations and our synthetic data. 
The trend of the velocity deviations observed in the MWC~758 disk follow closely with those in our synthetic data of an eccentric disk with an eccentricity of 0.1 at a radius of 0\farcs32, especially at azimuthal angles larger than $180\arcdeg$. The main difference lies in those at azimuthal angles around 135$\arcdeg$, which could be due to an inflow within the inner cavity toward the central star.
Thus, our imaging simulations demonstrate that with the same data quality as the observations, 
our analysis is able to detect the velocity deviations due to the eccentric gas motion with an eccentricity of 0.1 and correctly constrain the eccentricity and the position angle of the pericenter. 


\subsection{Dust-Gas Coupling and Planet's Orbit}

Our constraint on the eccentricity and pericenter of the gas rotation in the MWC~758 disk could suggest the presence of hypothetical planetary companion(s) of several $M_J$, interior to the outer edge of the eccentric dust cavity \citep{Kley_2006}.
Several studies have proposed an interior planet to explain the observed spiral arms and dust clumps in the MWC~758 disk \citep{Baruteau_2019, Calcino_2020}, which could also support the planetary origins of our observed eccentricity in the gas rotation.
\par

\citet{Hsieh_2012} provides an analytical steady-state solution to the secular relationship between the dust and gas orbital phase and eccentricity as, 
\begin{align}
    E_d & =E_f\frac{1}{\sqrt{1+\sfrac{1}{\tau^2_{s,sec}}}}\exp[i\arctan(\frac{-1}{\tau_{s,sec}})]\nonumber \\
    & +E_g\frac{1}{\sqrt{1+\tau^2_{s,sec}}}\exp[i\arctan(\tau_{s,sec})], 
    \label{eq:ef}
\end{align}
where $E_d$, $E_g$, and $E_f$ are the complex eccentricities of the dust, gas, and planet's orbits, respectively, to include the information of both the eccentricity magnitude and pericenter phase angle\footnote{Specifically, $E=e \exp{i \omega}$ with the subscript $d$, $g$, or $f$. Note that $E_f$ is the forced eccentricity driven by the planet's eccentricity $E_p$. $e_f \lesssim e_p$ in the vicinity of the planet.}.
$\tau_{s,sec}$ is the dimensionless secular stopping time, defined as the ratio of the stopping time to the precession time scale (i.e., the secular time) of a dust particle,   
and $\tau_{s,sec}$ is smaller than the Stokes number.
Thus, our results showing the consistent eccentricity of the gas and dust at the radius of the inner cavity in the MWC~758 disk could suggest that the dust and gas close to the inner cavity are sufficiently coupled on the secular timescale, so that $\tau_{s,sec} < 1$ \citep{Hsieh_2012}.

\par

Applying the secular perturbation theory for dust motion with the gas motion from hydrodynamical simulations, \citet{Hsieh_2012} showed that for transitional disks with lower gas surface densities, as in the case of MWC~758 (see Table \ref{tab:param}), the dust grains with sizes smaller than 0.1~mm exhibit an eccentricity tightly coupled to the gas eccentricity. On the other hand, the dust with grain sizes of $\gtrsim$ 1~cm could show a smaller eccentricity than the gas inside the inner cavity, while the dust and gas eccentricities could remain the same at outer radii, where the gas surface density is higher. 
The ALMA 0.88~mm and VLA 9~mm observations show two bright clumps in the MWC~758 disk \citep{Marino_2015, Dong_2018, Casassus_2019}. The presence of these dust clumps has been interpreted as dust trapping by vortices, where the larger dust grains are moderately coupled to the gas and drift to the center of the pressure maximum \citep{Casassus_2019, Baruteau_2019}. 
These ALMA and VLA results suggest that in the MWC~758 disk, the cm-sized dust grains are present, and part of the dust is weakly coupled to the gas, implying that the Stokes number is on the order of unity or slightly larger. Nevertheless, $\tau_{s,sec}$ is expected to be smaller than the Stokes number by two orders of magnitude \citep{Hsieh_2012}, and the ALMA 0.88~mm continuum observations are more sensitive to the distribution of the mm-sized dust grains. 
Therefore, our estimated gas eccentricity identical to the eccentricity of the dust ring observed in the 0.88~mm continuum is consistent with the theoretical expectation as demonstrated in the secular theory and numerical simulations for dust grains with sizes of 0.1~mm and 1~cm in \citet{Hsieh_2012}.

\par

Equation \ref{eq:ef} also shows that if a planet is on an eccentric orbit, the dust and gas orbits could have different eccentricites and position angles of their pericenters. 
\citet{Calcino_2020} demonstrate that the interaction with a planet having an orbital eccentricity of $\sim0.4$ can reproduce the complex morphology of the MWC~758 disk, including the double spiral arms, dust arc, inner cavity, and the twisted iso-velocity contours in the $^{13}\mathrm{CO}$ lines, in hydrodynamic simulations.
Our results showing the consistent dust and gas eccentricity at the radius of the inner cavity in the MWC~758 disk could suggest that the planet in the MWC~758 disk is on a circular orbit, or solely due to a very small $\tau_{s,sec}$, as the latter is indicated by the similar gas and dust eccentricities. 
Although our results also show the orientations of the dust and gas eccentric orbits are consistent within the uncertainties, the uncertainty in the position angle of the pericenter of the gas eccentric orbit is large with the current data. 
Thus, it remains unclear if the dust and gas eccentric orbits have a phase lag. 
Future observations at higher resolution and sensitivity are needed to better measure the orientation of the gas eccentric orbit, which will put an additional constraint on the planet's orbit in the MWC~758 disk. 

\par


\section{Summary}\label{sec:summary}
MWC~758 is so far the only known protoplanetary disk with confirmed dust eccentricity. To constrain its gas eccentricity, we analyzed the ALMA archival data of the $^{13}\mathrm{CO}$ and $\mathrm{C}^{18}\mathrm{O}$ emission in the MWC~758 disk. We fitted the data with a Keplerian disk model and found the velocity deviations from the circular Keplerian rotation around the inner cavity at a radius of 0\farcs32. The velocity deviations at a radius of 0\farcs32 are redshifted (or positive) at most azimuthal angles, especially along the major and minor axes of the disk, and the magnitude is $\sim$10\% of the local Keplerian velocity. The velocity deviation is significantly negative only at azimthual angles around 135$\arcdeg$ from the minor axis in the northwest.

We discussed various possible origins of the velocity deviations, including the pressure gradient, height of the emitting layer, infall motion, inner warp, and eccentricity in the disk, and calculated the expected velocity deviations of these mechanisms to compare with the observations. The velocity deviations due to the pressure gradient, the height of the emitting layer, and the warp  are expected to show sinusoidal profiles with both positive and negative deviations, and thus these mechanisms cannot explain the overall trend of the observed velocity deviations. An eccentric orbital motion in the MWC~758 disk is expected to induce positive velocity deviations at all azimuthal angles, which best explains the overall trend of the observed velocity deviations, while the negative velocity deviations at azimthual angles around 135$\arcdeg$ could be explained with an infalling flow. Thus, our results suggest that the non-Keplerian rotation close to the inner cavity in the MWC~758 disk could be caused by the combination of an eccentric orbital motion and an infalling flow. 

Assuming that the positive velocity deviations at $r = 0\farcs32$ are due to the eccentric gas motion, we measured the eccentricity to be $0.1\pm0.04$ and the position angle of the pericenter to be $270\arcdeg\pm32\arcdeg$ from the azimuthal profiles of the velocity deviation. In addition, there is no detectable eccentricity at radii larger than 0\farcs45 in the MWC~758 disk. The measured eccentricity and position angle of the pericenter of the gas rotation are consistent with those of the dust ring around the inner cavity at the radius of 0\farcs32, $0.1\pm0.01$ and $275\arcdeg\pm10\arcdeg$, respectively. 

Our results suggest the presence of a gas giant planet in the inner cavity in the MWC~758 disk, which excites the disk eccentricity. Furthermore, the consistent eccentricity of the gas and dust around the inner cavity in the MWC~758 disk hints that the dust and gas there are secularly well coupled, so the two components follow the same eccentric orbital motion. This is consistent with the expectation from the VLA 9~mm observations showing that there is no significant trapping of larger dust grains close to the inner cavity.

It has also been proposed that the possible planet in the inner cavity is on an eccentric orbit, which may cause an offset in the position angles of the pericenters of the dust and gas orbital motions. Nevertheless, our constrained position angle of the pericenter of the eccentric orbital motion of the gas is not accurate enough to make a meaningful comparison with that of the dust due to the limited resolution and senstivity of the data. Future observations better resolving the azimuthal profile of the velocity deviation from the Keplerian rotation could put an additional constraint on the planet's orbit in the MWC~758 disk.

\acknowledgments
We thank St\'ephane Guilloteau, Anne Dutrey, Edwige Chapillon, Ya-Wen Tang, Patrick Koch, and Bo-Ting Shen for discussions on our observational results and their help with fitting the molecular-line data with Keplerian disk models using DiskFit. We thank He-Feng Hsieh for discussions on our numerical simulations using FARGO3D and for his help with the additional treatment for the grain size in our simulations.
This paper makes use of the following ALMA data: ADS/JAO.ALMA\#2012.1.00725.S. ALMA is a partnership of ESO (representing its member states), NSF (USA) and NINS (Japan), together with NRC (Canada), MOST and ASIAA (Taiwan), and KASI (Republic of Korea), in cooperation with the Republic of Chile. The Joint ALMA Observatory is operated by ESO, AUI/NRAO and NAOJ. H.-W.Y.\ acknowledges support from the Ministry of Science and Technology (MOST) in Taiwan through grant MOST 108-2112-M-001-003-MY2 and MOST 110-2628-M-001-003-MY3 and from the Academia Sinica Career Development Award (AS-CDA-111-M03). P.-G.G.\ acknowledges support from MOST in Taiwan through grant MOST 109-2112-M001-052. 

\bibliography{main}
\bibliographystyle{aasjournal}

\begin{appendix}

\section{Degeneracy between the inclination and the stellar mass}\label{deg_i-Ms}
\begin{figure*}[ht!]
\epsscale{1}
\plotone{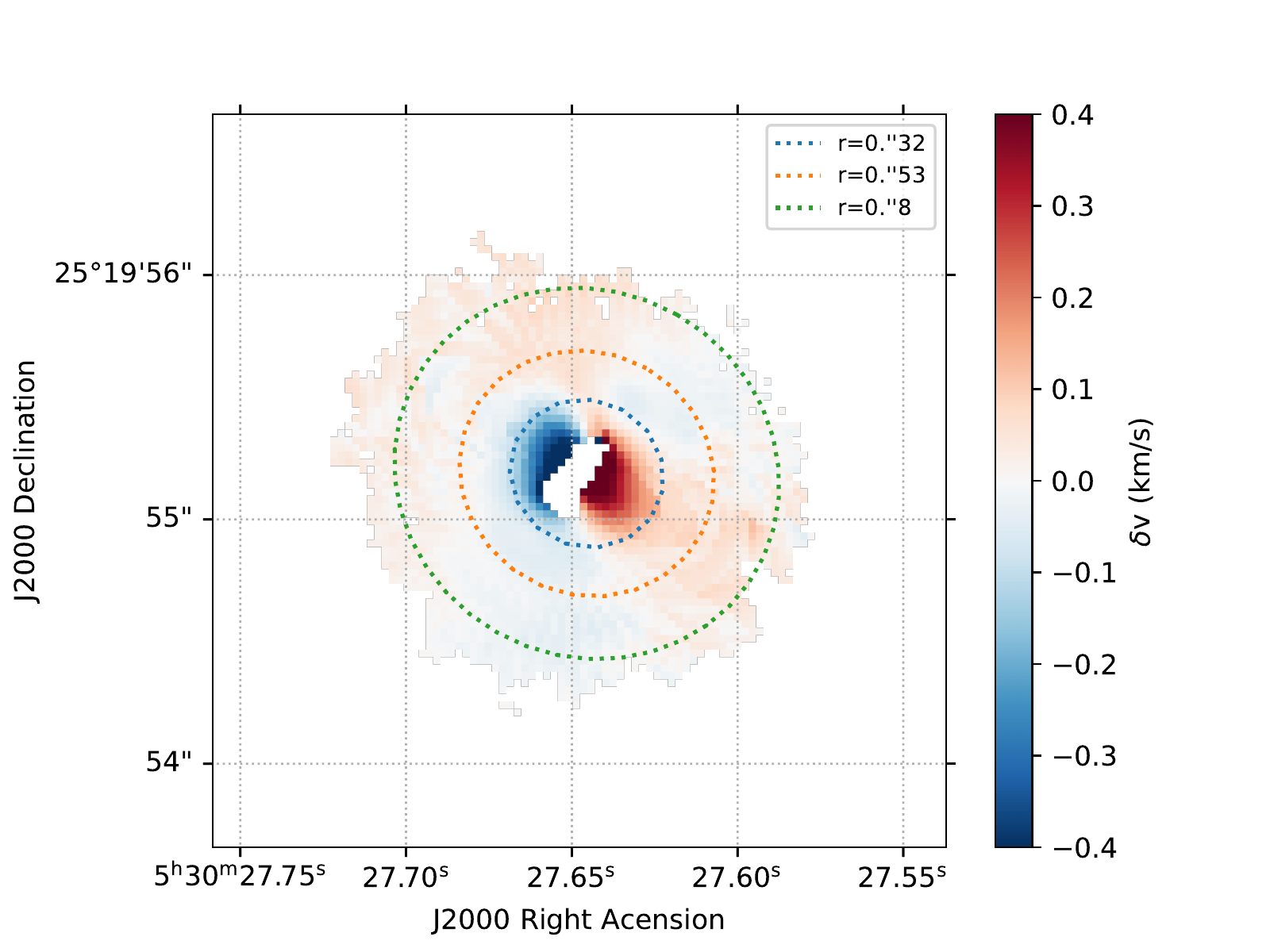}
\caption{Difference between the projected velocity maps of the two best-fit Keplerian disk models with inclination fixed at 21$\arcdeg$ and with the stellar mass fixed at 2~$M_\sun$.
\label{fig:mod_compare}}
\end{figure*}

We found that with our current data, the inclination and stellar mass of MWC~758 could not be constrained simultaneously. Thus, in this paper, we present two best-fit models, one with the inclination fixed at 21$\arcdeg$ and one with the stellar mass fixed at 2~$M_\sun$ (Section~\ref{subsec:circmodel}). We compared the projected velocity maps of these two models by subtracting one from the other. Figure~\ref{fig:mod_compare} shows the difference between the two velocity maps. At radii larger than 0\farcs3, the velocity differences between these two models are typically smaller than the observational error bar and the velocity channel width of 0.1 km/s. Although the velocity residuals can become larger than the observational error bar at radii smaller than 0\farcs3, this inner central region is smaller than the inner cavity in the disk and is not the main focus of the present paper. The two models have consistent projected velocity maps within the uncertainty in the regions of our interest. Thus, the choice of the best-fit models does not affect our subsequent analysis of the velocity deviation from the Keplerian rotation because the analysis is based on the comparison between the projected velocity maps from the model and the observations, and this has also been tested in Appendix~\ref{ecc_i16}. We also applied our eccentric model (Section \ref{subsec:ecc}) to this velocity difference map, and our analysis indeed did not result in any false detection of non-zero eccentricity.

\section{$\mathrm{C}^{18}\mathrm{O}$ $J=3-2$ Velocity Maps}\label{c18o_fitting}
\begin{figure*}[ht!]
\epsscale{1.15}
\plotone{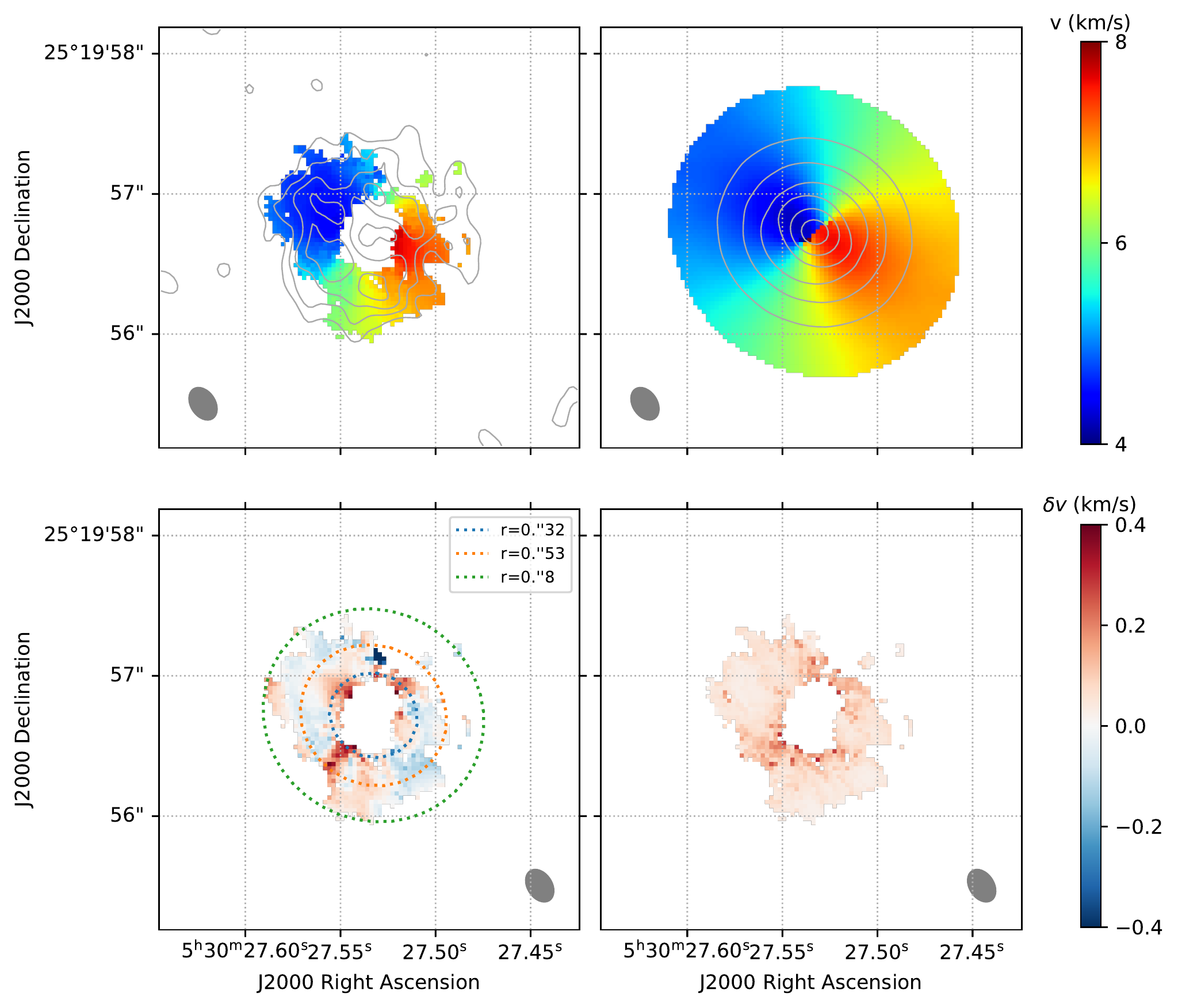}
\caption{Same as Figure \ref{fig:vel_plots}, but with $\mathrm{C}^{18}\mathrm{O}$ $J=3-2$ \label{fig:vel_plots18}}
\end{figure*}

Figure \ref{fig:vel_plots18} shows the observed velocity map (upper left), the $i=21{\arcdeg}$ $M_{\star}=1.2 M_{\odot}$ Keplerian model velocity map (upper right), the residuals after subtracting the model velocity map from the observed velocity map (lower left), and the uncertainty in the velocity (lower right) of the C$^{18}$O $J=3-2$ emission.
The best-fit parameters of our Keplerian disk models to the C$^{18}$O data are listed in Table~\ref{tab:c18o_param}.
\par
The $\mathrm{C}^{18}\mathrm{O}$ $J=3-2$ velocity residuals are positive around nearly the full azimuthal extent at $r=0\farcs32$, but the S/N is below $2\sigma$ and too low for a robust analysis of the origins of the velocity residuals. We can also see negative residuals at outer radii, similar to the trend in the velocity residuals of $^{13}\mathrm{CO}$ $J=3-2$.

\begin{deluxetable}{cchlDlccc}[ht!]
\tablecaption{Summary of $\mathrm{C}^{18}\mathrm{O}$  Best-Fit Model\label{tab:c18o_param}}
\tablewidth{0pt}
\tablehead{
\colhead{Parameter (units)} & \colhead{Fixed Inclination} & \nocolhead{} & \colhead{Fixed Stellar Mass} }
\startdata
PA ($\arcdeg$) & $61.4\pm0.5$ &  & $61.3\pm0.5$
\\
inclination ($\arcdeg$) & $-21.00$ &  & $-16.7\pm0.1$
\\
$v_{\rm sys}$ (km~s$^{-1}$) & $5.90\pm0.02$ &  & $5.89\pm0.07$
\\
$M_{\star}$ ($M_{\odot}$) & $1.35\pm0.02$ &  & $2.00$
\\
$r_{in}$ (au) & $39.3\pm4.0$ &  & $38.9\pm3.6$
\\
$r_{out}$ (au) & $157.7\pm3.8$ &  & $156.8\pm9.4$
\\
$\log\Sigma_0$ (cm$^{-3}$)& $15.9\pm0.1$ &  & $15.8\pm0.1$
\\
$pp$  & $4.7\pm0.7$ &  & $4.4\pm0.5$
\\
$T_0$ (K) & $24.4\pm4.7$ &  & $24.2\pm3.4$
\\
$qq$ & $0.33\pm0.22$ &  & $0.50\pm1.2$
\\
$H_0$ (au) & $18.1\pm13.6$ &  & $16.0\pm23.7$
\\
$hh$ & $-0.6\pm14.4$ &  & $-1.3\pm13.5$
\enddata
\tablecomments{PA is the position angle of the major axis of the disk. Inclination is defined as the angle between the normal axis of the disk and the line of sight. $v_{\rm sys}$ is the systemic velocity. Power-law profiles of the disk properties are parameterized as: (1) column density $\Sigma(r)=\Sigma_0(\frac{r}{r_0})^{-pp}$, (2) temperature $T(r)=T_0(\frac{r}{r_0})^{-qq}$, and (3) scale height}
\end{deluxetable}

\section{Analysis with the disk model with the inclination of 16$\arcdeg$}\label{ecc_i16}
\begin{figure*}[ht!]
\epsscale{1.2}
\plotone{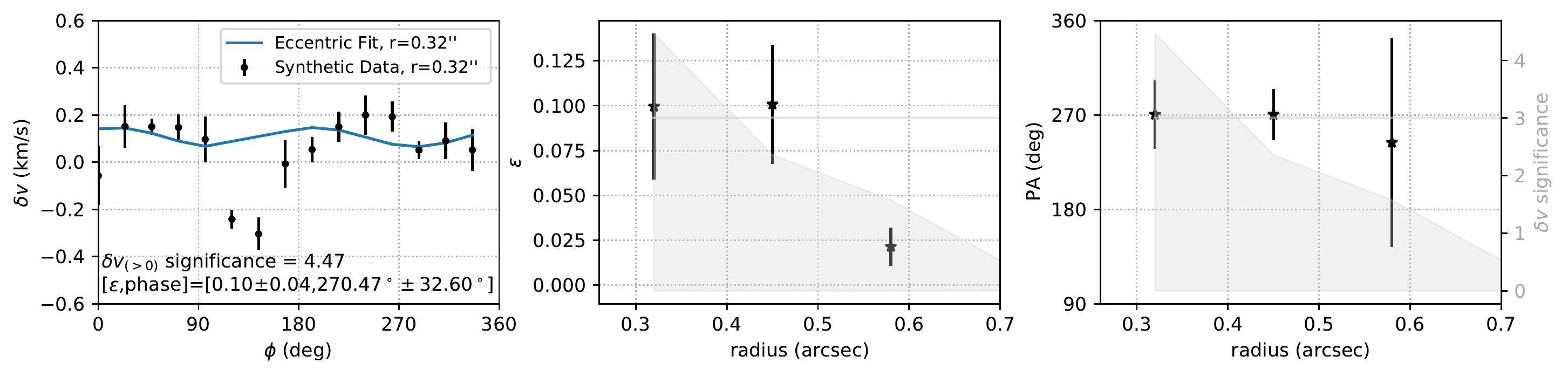}
\caption{Same as Fig.~\ref{fig:13COecc} but the analysis was performed with the best-fit Keplerian disk model with the inclination of 16$\arcdeg$ and $M_\star$ of 2~$M_\sun$.
\label{fig:13COecc_i16}}
\end{figure*}

We have also performed the same analysis described in Section \ref{subsec:ecc} with our best-fit Keplerian disk model with the inclination of 16$\arcdeg$ and $M_\star$ of 2~$M_\sun$. The azimuthal profile of the extracted velocity deviations from the circular Keplerian rotation at $r=0\farcs32$ is shown in Figure \ref{fig:13COecc_i16}. We fitted the eccentric model to the positive velocity deviations. The eccentricity and the position angle of the pericenter were estimated to be $0.1\pm0.04$ and $270\arcdeg\pm33\arcdeg$ at $r=0\farcs32$. The radial profiles of the estimated eccentricity and position angle of the pericenter in the MWC~758 disk are shown in the middle and right panels in Figure \ref{fig:13COecc_i16}, respectively. These results are consistent with those estimated with the Keplerian disk model with the inclination of 21$\arcdeg$ and $M_\star$ of 1.2~$M_\sun$ within the uncertainties.  

\end{appendix}

\end{document}